\documentclass[twocolumn]{article}
\usepackage[cm]{fullpage}
\usepackage{natbib}
\usepackage{graphicx}
\usepackage{url}
\usepackage{doi}
\usepackage{amsmath, amssymb}
\usepackage{abbrevs}
\usepackage{hyperref}
\usepackage{amsmath}
\renewcommand{\mod}[1]{\lvert #1 \rvert}

\newcommand{\ct}{\mathit{ct}}
\newcommand{\ET}{\mathbf{ET}}
\newcommand{\Uz}{z}
\newcommand{\fz}{\zeta}
\newcommand{\et}{\mathit{et}}
\newcommand{\asin}{\mbox{asin}}

\newcommand{\median}{\mbox{median}}
\newcommand{\MAD}{\mbox{MAD}}
\newcommand{\WSS}{\overline{\mathrm{WSS}}}

\newabbrev\rtqPCR{reverse-transcriptase qPCR (rt-qPCR)}[RT-qPCR]
\newabbrev\qPCR{quantitative polymerase chain reactions (qPCR)}[qPCR]
\begin{document}
\title{Data Exploration, Quality Control and Testing in Single-Cell qPCR-Based Gene Expression Experiments}
\author{Andrew McDavid\,$^{1,2}$, Greg Finak\,$^{2}$, Pratip K. Chattopadyay\,$^3$, Maria Dominguez\,$^3$,\\
 Laurie Lamoreaux\,$^4$, Steven S. Ma\,$^4$, Mario Roederer\,$^3$ and Raphael Gottardo\,$^{1,2}$\footnote{to whom correspondence should be addressed} \\
\\
$^{1}$Department of Statistics, University of Washington, Seattle, WA\\
$^{2}$Vaccine and Infectious Disease Division, Fred Hutchinson Cancer Research Center, Seattle, WA \\
$^{3}$ImmunoTechnology Section, Vaccine Research Center, NIAID, NIH,
Bethesda, MD \\
$^{4}$Immunology Laboratory, Vaccine Research Center, NIAID, NIH, Bethesda, MD }

\maketitle
\begin{abstract}
\noindent\textbf{Motivation:}
Cell populations are never truly homogeneous; individual cells exist in biochemical states that define functional differences between them.
%Measurement at the single-cell level provides important information about cell population heterogeneity that would otherwise be masked by ``bulk'' assays.
New technology based on microfluidic arrays combined with multiplexed \qPCR now enables high-throughput single-cell gene expression measurement, allowing assessment of cellular heterogeneity.
However very little analytic tools have been developed specifically for the statistical and analytical challenges of single-cell \qPCR data.

\noindent \textbf{Results:}
We present a statistical framework for the exploration, quality control, and analysis of single-cell gene expression data from microfluidic arrays.
We assess accuracy and within-sample heterogeneity of single-cell expression and develop quality control criteria to filter unreliable cell measurements.
% Three independent data sets measured gene expression at either the single-cell or hundred-cell (bulk) level.
We propose a statistical model accounting for the fact that genes at the single-cell level can be \textit{on} (and for which a continuous expression measure is recorded) or dichotomously \textit{off} (and the recorded expression is zero).
Based on this model, we derive a combined likelihood-ratio test for differential expression that incorporates both the discrete and continuous components.
Using an experiment that examines treatment-specific changes in expression, we show that this combined test is more powerful than either the continuous or dichotomous component in isolation, or a $t$-test on the zero-inflated data.
While developed for measurements from a specific platform (Fluidigm), these tools are generalizable to other multi-parametric measures over large numbers of events.

\noindent \textbf{Availability:}
All results presented here were obtained using the
\texttt{SingleCellAssay} R package available on
\href{http://github.com/RGLab/SingleCellAssay}{GitHub} (\url{http://github.com/RGLab/SingleCellAssay}).

\noindent \textbf{Contact:} \href{rgottard@fhcrc.org}{rgottard@fhcrc.org}

\noindent \textbf{Supplementary Material:}
Supplementary data are available.
\end{abstract}

\section{Introduction}
The development fluorescence-based flow cytometry (FCM) revolutionized single-cell analysis.
Although even nominally-homogeneous populations of cells sorted by flow cytometry using established surface markers may appear monolithic, mRNA expression of specific genes within these cells can be heterogeneous \citep{Dalerba2011Singlecell} and could further discriminate cell subsets.
On the other hand, classical gene expression experiments (micro-arrays, RNA-seq, qPCR) richly characterize a cellular population, but at the cost of reporting a summation of expression from many individual cells.
Recent advances in microfluidic technology now permit performing thousands of PCRs in a single device, enabling rich gene expression measurements at the single-cell level across hundreds of cells and genes \citep{Kalisky2011}.
This provides a technology that probes the stochastic nature of biochemical processes, resulting in relatively large cell-to-cell expression variability.

This heterogeneity may carry important information: thus single cell expression data should not be analyzed in the same fashion as population-level data.
At the scale of a single cell, biological variability (the object of interest) and technical variability (a nuisance factor) are often of the same magnitude, making it difficult to distinguish between the two.
The dichotomous nature of single-cell data further complicates matters: measurements on individual cells may be dichotomously absent due to real biological effects, so are not always present on a continuum.
These features of single-cell data require special attention during analysis.

Here we focus on the \rtqPCR-based Fluidigm (San Francisco, CA)  single-cell gene expression assay, which provides simultaneous measurements of up to 96 genes on mRNA sources as minute as a single cell.
In traditional \rtqPCR, despite careful measurement of starting concentrations of cDNA, correction for differences in quantities of starting material below the limit of detection is necessary for reliable results  \citep{Vandesompele2002Accurate}.
Subtraction of internal control genes, or averages thereof is typically used (\textit{e.g.}, the $\Delta$-Ct method), and results are often reported in numbers of copies, or fold increase per cell~\citep{Schmittgen:2008tt}.
In array-based gene expression, differences in hybridization and washing of non-specific DNA between chips require additional correction.

Such normalization schemes are not directly applicable in single-cell gene expression experiments, nor is it obvious that they are needed.
For single cells, the individual cell is the atomic unit of normalization and the amount of starting material naturally measured in number cells per reaction.
Even if one attempted direct application of traditional normalization approaches, the dichotomous nature of single-cell expression hinders their use.

Nonetheless, it is important to test for and address any technical biases.
We present a filtering approach for removing outlying measurements at the single-cell level that accounts for the dichotomous nature of the data. Using concordance measures derived from three data sets where gene expression was measured at the single-cell and hundred-cell levels, we show that classical \rtqPCR type normalization is not necessary with single-cell multiplexed PCR data and that our filtering step removes technical artifacts that most severely impact quantitation.

A typical goal of gene expression experiments is to search for differential expression across groups.
The dichotomous nature of expression in Fluidigm introduces problems for testing differential representation of cell subsets characterized by expression patterns, as well.
Traditional tests of differential expression such as the t-test or other approaches based on normality are likely inappropriate for zero-inflated data \citep{Smyth2004Linear,Gottado2006a}.
Approaches to this problem have varied. \citet{Powell2012Single} used a winsorized z-transformation of the expression values, then treated them as continuous.
\citet{Glotzbach2011Information} used the non-parametric, Kolmorgov-Smirnov test for differences in distribution to find differentially expressed genes, after winsorizing.
\citet{flatz_2011} dichotomized the expression and worked with the binary trait.
 However, as we will see later, both the continuous and discrete parts of the measurements are informative for differential expression and should be used.
A parametric test allows directions of difference to be assessed.

Here we propose a discrete/continuous model for single-cell expression data based on a mixture of a point mass at zero and a log-normal distribution.
Using this model, we derive a likelihood ratio test that can simultaneously test for changes in mean expression (conditional on the gene being expressed) and in the percentage of expressed cells.
\section{Methods}
\subsection{Data sets and notations}
\label{sec:datasets}
We use three Fluidigm single-cell gene expression data sets described below.
We offer a brief overview of the assay technology used for our data.
Desired cells (\textit{e.g.} antigen-specific CD8+T cells) are selected and lysed, and a cDNA library is generated through \rtqPCR.  A short ( c. 15 cycle), multiplexed pre-amplification selects and enriches for the desired genes.
These products are loaded onto the Fluidigm chip and gene-specific primers are added for single-cell gene expression quantification.
For the data presented here we used a $96\times 96$ format plate, \textit{i.e.} 96 genes across 96 cells.
The design of the chip generates each combination of the 96 genes and 96 enriched cDNA libraries producing 9216 separate PCR reactions.
After each cycle, the fluorescence is read. The cycle (or interpolated fraction thereof) at which the fluorescence crosses a pre-determined threshold is recorded, defined as the ``\textit{ct}'' value. For all data sets considered here, primers were chosen to have $>90\%$ amplification efficiency. %cite Maria's article

\noindent\textbf{Data set A:}  Twenty-eight $96 \times 96$ format plates of CMV- or HIV-specific CD8+ single cell T cells were isolated from 16 individuals.
The donors' cells were stimulated with one of four tetramers.
Cells were sorted immediately after tetramer incubation (``unstimulated'') or after 3 hours of exposure (``stimulated'').
Approximately 90 individual cells were measured for each patient-stimulation combination (``unit'').

%In contrast, in data sets B and C, no individual-level treatments were undertaken.
\noindent\textbf{Data set B:} Ten subjects were considered, and approximately 180 cells were sorted per subject, with each subject crossed between two arrays.

\noindent\textbf{Data set C:} Two subjects were considered. Fluorescent staining of CD4+ T cells allowed cytometric sorting into CD154+/- sub-populations.
Approximately 40 cells were sorted per sub-population  per subject  across three arrays.

Additionally, for each individual and treatment within each data set,  aggregates of 100 cells (\textit{i.e.}, 100 cells per well on the array) were isolated and assayed by Fluidigm technology.
The expression measured in these 100-cell aggregates, after dividing by 100, provides a ``biological'' average of expression per--cell, and can be compared to an \emph{in silico} average of the single-cell measurements.
The \textit{concordance} between these two averages serves as a measure of experimental fidelity~\citep{Lin1989Concordance}.
\begin{figure}[!tbp]
  \centering
\includegraphics[width=.5\textwidth]{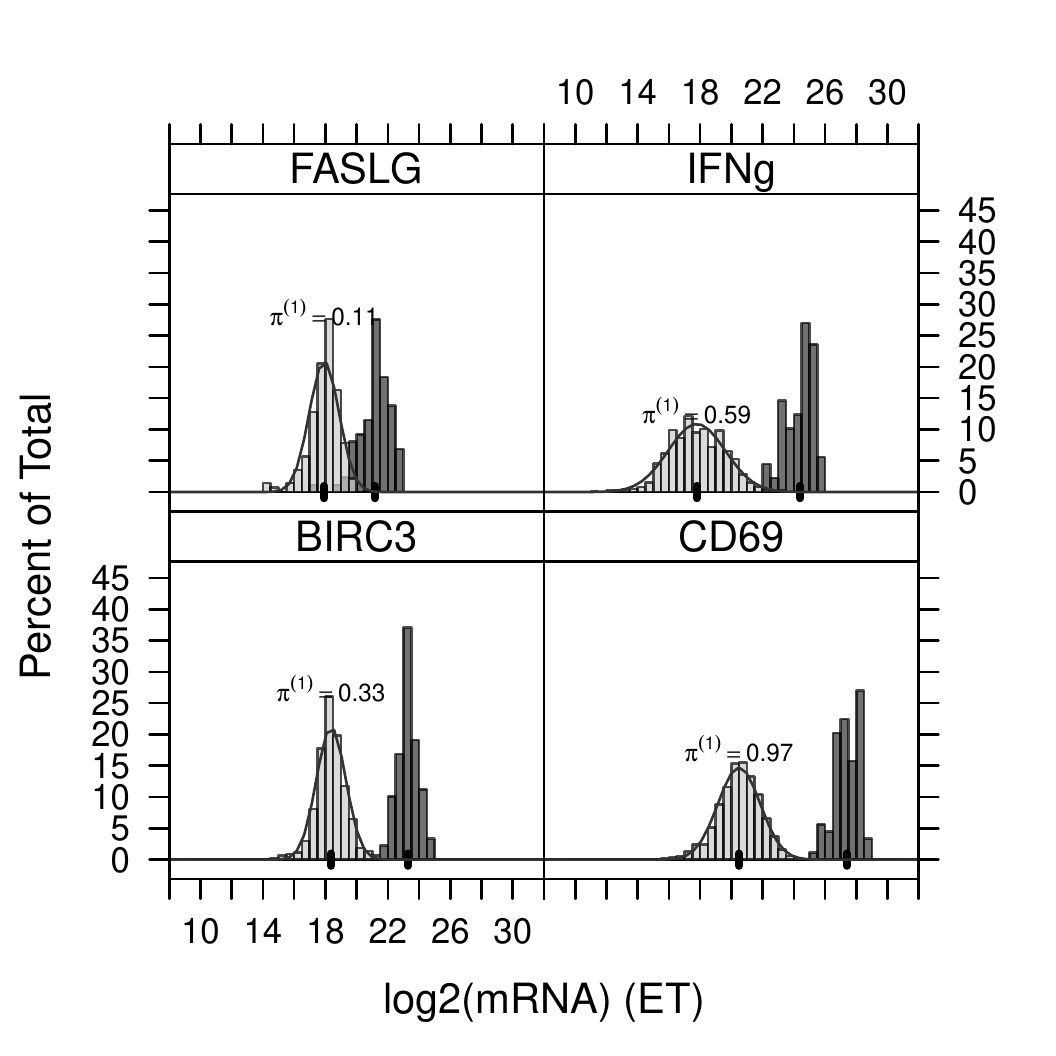}
  \caption{Histogram and theoretical (normal) distribution of $(\et_{ij} | v_{ij}=1)$ for single cell (left, light gray) and hundred cell experiments (right, dark gray).
Genes FASLG, IFN-$\gamma$, BIRC3 and CD69 are depicted.
The frequency expression of each gene in the single cell experiments $\pi^{(1)}$ is printed above each histogram.
The mean of the hundred cell and single cell experiments is indicated by a thick black line along the x-axis.}
  \label{fig:hist}
\end{figure}
\noindent\textbf{Notations:}
The standard assumptions of qPCR-based assays apply to the Fluidigm
technology, namely that the cycle threshold ($\ct$) is inversely
proportional to the log of fluorescence.
The fluoresence is directly proportional to the starting concentration of mRNA~\citep{Higuchi:1992ud,Karlen:2007bd}.
The Fluidigm instrument returns the cycle threshold ($\ct$), however, we find it more useful to work with the complement of $\ct$, which we define as the \emph{expression threshold} ($et$)
\begin{equation*}
  \et = c_{\mathrm{max}} - \ct
\end{equation*}
where $c_{\mathrm{max}}$ is the maximum number of cycles used, 40 in our case. 
Assuming all reactions are in the exponential amplification phase, this quantity should be directly proportional to the
log-abundance of mRNA, plus an intercept term corresponding to the number of cycles it takes for the minimally-detectable quantity of mRNA to cross threshold.
If the fluorescence does not cross the threshold after 40 cycles, then the Fluidigm instrument records a value of N/A, and we say that the gene is \emph{not detected}.
As we will see in the results section, detected genes typically have a value of $\ct$ much less than $c_{\mathrm{max}}$ suggesting that undetected genes might be regarded as unexpressed genes. This assumption is supported by the idea that transcription of mRNA is thought to occur in bursts of activity \citep{Levsky2002SingleCell, Kaufmann2007Stochastic}, followed by quiescence.
Other authors have noted this feature in single cell expression studies as well \citep{Glotzbach2011Information}.
When looking at the concordance of the single-cell and hundred-cell experiments, this assumption is reasonable and leads to better concordance than omitting the N/A values.
As a consequence, we treat the undetected genes as unexpressed genes, and we set the corresponding $\et$ value to $-\infty$, so that the mRNA abundance is zero (\textit{i.e.} $2^{\et} = 0$).
For a fixed sample or experimental unit, let us denote by $\et_{ij}$ the expression threshold of \emph{well} $i$ and \emph{gene} $j$, for $i=1, \dotsc, I$ and $j=1, \dotsc, J$. This results in a matrix of $\log_2$ based expression values, $\ET=(\et_{ij})$, just as in array-based gene expression. Similarly, we will denote by $\mathbf{Y}=(y_{ij})$ the matrix of untransformed expression values where $y_{ij}=2^{\et_{ij}}$.
Usually a well contains one cell but the Fluidigm technology can be used with multiple cells per well to quantify the gene expression of a mixture of cells. As a consequence, we prefer to use the term ``well'' instead of ``cell''.
In the three data sets used here, wells will contain either one or hundred cells. Finally, several biological units are typically measured in an experiment, and in this case we will use an extra index $k$ to refer to the biological units.
\subsection{A model for single cell expression}
As described previously, for a given cell, a gene can be defined as \textit{on} (\textit{i.e.} a positive $\et$ value is recorded) or as \textit{off} (\textit{i.e.} the gene is undetected and $y_{ij}=0$). To simplify our model, we will denote by $v_{ij}=\mathbf{1} \left[ y_{ij}=0 \right]$ the indicator variable equal to one if the gene $j$ is expressed in well $i$ and zero otherwise. Following classical statistical conventions, we use upper cases to denote the random variables, and lower cases to denote the values taken by these random variables. Using these notations, we introduce the following model of single-cell expression
\begin{eqnarray}
\label{eq:etdist}
(Y_{ij}|V_{ij}=1)&\sim&\mathrm{logNormal}(\mu_j, \sigma_j^2)\\
(Y_{ij}|V_{ij}=0)&\sim&\delta_0\\
V_{ij}&\sim& \mathrm{Be}(\pi_j)
\end{eqnarray}
where $\delta_0$ denotes a point mass at zero,  $\mu_j$ and $\sigma_j^2$ are the $\log_2$-based mean and variance expression level parameters conditional on the gene being expressed (\textit{i.e.} $V_{ij}=1$), and $\pi_j$ is the frequency of expression of gene $j$ across all cells. In the data sets considered here, the frequency of expression greatly varies across genes from 0 to .99 with a median value of $\pi_j$ around .1 (see Supplementary Figure 1).
Note that assuming a log-Normal model for $(Y_{ij}|V_{ij}=1)$ is equivalent to modeling $(\ET_{ij}|V_{ij}=1)$ as normally distributed.
The empirical distribution of the data (Figure \ref{fig:hist} and Supplementary Figures 4-6) motivates our selection of a log-normal distribution and follows observations of previous authors \citep{Bengtsson2005Gene}.

Thus in a particular gene, three parameters characterize the expression distribution: $\mu_j, \sigma_j$, the mean and standard deviation of the $\et_{ij}|V_{ij}=1$,  and $\pi_j$, the Bernoulli probability of expression.

% \begin{table}[!t]
% \processtable{This is table caption\label{Tab:01}}
% {\begin{tabular}{llll}\toprule
% head1 & head2 & head3 & head4\\\midrule
% row1 & row1 & row1 & row1\\
% row2 & row2 & row2 & row2\\
% row3 & row3 & row3 & row3\\
% row4 & row4 & row4 & row4\\\botrule
% \end{tabular}}{This is a footnote}
% \end{table}

\subsection{Quality Control and Filtering}
\label{sec:qual-contr-filt}
The Fluidigm assay is sensitive, and due to the exponential amplification of starting mRNA, even minute contamination can render a measurement unreliable.
Similarly, variation in cell preparation can have significant impact on the resulting experiment and data, such as unintentional empty wells, which would distort estimates of $\pi_j$. This suggests testing for, and possibly removing outliers before conducting further analysis.
We examine both the discrete component $v_{ij}$ and the continuous component $(et_{ij}|v_{ij}=1)$ in screening for outliers.
We define the robust z-transformed positive expression value as
\[
\Uz_{ij} \equiv \frac{\et_{ij} - \median_i (\et_{i j})}{k \cdot \MAD_i(\et_{i j})},
\label{eq:etz}
\]
where the median and median absolute deviation are calculated, for a given gene, over expressed cells (\textit{i.e.} $v_{ij}=1$), and $k=1.48$ is a scaling constant that gives the standard deviation in terms of the MAD for the normal distribution.
Next, let $f_i = \asin \sqrt {\bar{v}_{i\cdot}}$ be the Bernoulli variance-stabilizing transformation of the proportion of genes expressed in well $i$.
Then we define the robust z-transformed fraction as
\[
\fz_i \equiv \frac{f_i - \mbox{median}_i(f_i)}{k \cdot \mbox{MAD}_i(f_{i})}
\]
where the median, MAD and $k$ are as defined previously. This leads to the following steps for filtering:
\begin{enumerate}
\item Remove \emph{null} cells with no detected genes, \textit{i.e.} $V_{ij}=0$, for all $j$.
\item Pick threshold for $\Uz$ filtering ($t_z$); threshold for $\fz$ filtering ($t_\zeta$).%; and number of violations of $t_z$ permitted per well (\emph{noutlier}).
\item Calculate $\Uz_{ij}$ and $\fz_i$
\item Remove wells in which %\emph{noutlier} or more
genes have $\mod{\Uz} > t_z$ OR $\mod{\fz}>t_\zeta$.
\end{enumerate}
Step 1 removes wells where no cells were loaded, and thus all measured
expression values are null. It is important to perform this step first
to prevent break-down in the median and MAD estimates for the $\fz$'s
in experiments with many amplification or flow cytometry
failures. Finally, step 4 removes unreliable wells that either have an
extreme proportion of expression or %too many
 extreme cell$\times$gene expression values.
In practice, we find that picking $t_z=9, t_\zeta=9$ % and
%$\emph{noutlier}=2$
 works well for the data sets we consider here, see
section~\ref{results} (Results).

\subsection{Testing for ET differences between experimental groups}
\label{sec:test-et-diff}
One typical goal of gene expression analysis is to test for difference in expression patterns between experimental units.
Here, we focus on testing differential gene expression between two paired-biological units, \textit{e.g.} before and after stimulation.
Our framework should be generalizable to other types of situations, see Discussion section.
The classical test for changes in mean for samples with continuous measurements is the $t$-test.
Conversely, if only a change in $\pi$ were of interest, then a contingency table test (Chi-square, Fisher's Exact or Bernoulli likelihood ratio) is appropriate.
However, in our case, we would like to test for a change in $\mu$ and $\pi$ simultaneously, since both could be biologically relevant.
Formally, we wish to test
\[
H_0: \pi_0 = \pi_1 \quad \mbox{and} \quad \mu_0 = \mu_1
\]
versus the alternative
\[
H_a: \pi_0 \neq \pi_1 \quad \mbox{and} \quad \mu_0 \neq \mu_1.
\]
This can be accomplished using a likelihood ratio test that would simultaneously test for differences in means or proportions of expression.

Suppose that $I$ wells are assayed in each unit, though the unbalanced case ($I_0\neq I_1$) would be treated similarly with obvious changes of notation.
Based on (\ref{eq:etdist}), the likelihood function for one gene across two biological units, omitting the gene index $j$ for clarity, is given by
\begin{equation}
\label{equ:likelihood}
\mathrm{L}(\boldsymbol{\theta}|\mathbf{y},\mathbf{v}) = \prod_k \pi_{k}^{n_{k}} (1-\pi_{k})^{I-n_{k}} \prod_{i\in_{S_{k}}}g(y_{ik}|\mu_{k},\sigma^2)
\end{equation}
where $\mathbf{y}$ and $\mathbf{v}$ are the vectors of observations for the gene across the two groups,
%$\mathbf{y}_j=\{y_{ijk}:i=1,\dots,I; k=1,2\}$,  $\mathbf{v}_j=\{v_{ijk}:i=1,\dots,I; k=1,2\}$,
$\boldsymbol{\theta}=\{\mu_{k},\sigma^2,\pi_{k}; k=0,1\}$ is the vector of unknown parameters, $S_k$ is the
set of cells expressing the gene in group $k$ (\textit{i.e.} $S_k=\{i: v_{ik}=1$\}) , $n_k=\sum_i v_{ik}$ is the number of cells expressing the gene in group $k$, and $g$ is the density function of the log-normal distribution with parameters $\mu_k$ and $\sigma_k^2$.
The likelihood ratio test (LRT) statistic $\Lambda(\mathbf{y},\mathbf{v})$ is then defined as the ratio of the null and alternative likelihoods obtained by replacing the unknown parameters with their null and alternative maximum likelihood estimates. Detailed derivations of the likelihood function and the LRT statistics are described in supplementary material.

An interesting observation is that the likelihood function given by (\ref{equ:likelihood}) is log-linear in the parameters $\pi$ and $(\mu, \sigma^2)$ since it is the product of the Bernoulli likelihood for all cells and the log-normal likelihood for the expressed cells.
It follows that the log-LRT statistic decomposes as a sum of a Bernoulli log-LRT test statistic and a lognormal log-LRT test statistic, since each component can be maximized independently. It thus combines the two sources of information in a natural way. In the results section we will show that our combined LRT test is more powerful than the Bernoulli or log-normal tests alone.

\section{Results}
\label{results}
\subsection{Distributional assumptions}
In Figure~\ref{fig:hist}, we observe good agreement between the empirical distributions of positive $\et$ values and their postulated normal distribution for four genes in data set A.
This confirms that a log-normal model for the positive expression level, $y_{ij}|v_{ij}=1$, is appropriate.
Even cells in the lowest quantiles of $\et$ (and lowest quantiles of expression) still have expression far away from the bound at 0, suggesting that undetected genes represent cells with null or negligible RNA abundance. It is also noteworthy that the difference between the means (shown as a heavy, vertical line) of the hundred cell replicates and single cell replicates is approximately $\log_2(100\cdot\pi^{(1)}_j)$ cycles, where $\pi_j^{(1)}$ is the expression frequency of gene $j$ in the single-cell experiments. As such, in genes with $\pi_1 \ll 1$, such as FASLG, this difference between means is smaller than genes with $\pi_j^{(1)} \approx 1$. As we will see the next section, inclusion of the unexpressed cells ($v_{ij}=0$) is important to accurately relate the expression level of the single-cell experiments to the hundred-cell experiments.

\subsection{Concordance between hundred-cell and single-cell experiments}
\label{sec:conc-betw-hundr}
The 100-cell aggregates (see section~\ref{sec:datasets}, \emph{data
  sets and notation}) allows us to assess the accuracy and reliability of our single-cell experiments by comparing this \textit{in-vitro} 100-cell expression to an \textit{in-silico} estimate obtained by averaging the expression of 100 single-cell measurements.
The \emph{in silico} average of signal in a gene $j$ and unit $k$ from 100 single-cell wells is $y_{jk}^{(1)} = \sum_{i=1}^{100} y_{ijk}/100$ where $y_{ijk}$ is the expression measurement of gene $j$ in cell $i$ and unit $k$. We compare this to the \emph{in-vitro} ``average'' of signal from a 100 cell aggregate. In this case, we just use the expression of a gene-unit and divide by the number of cells (100).

\begin{figure*}[htpb]
  \centerline{\includegraphics[height=.7\paperheight]{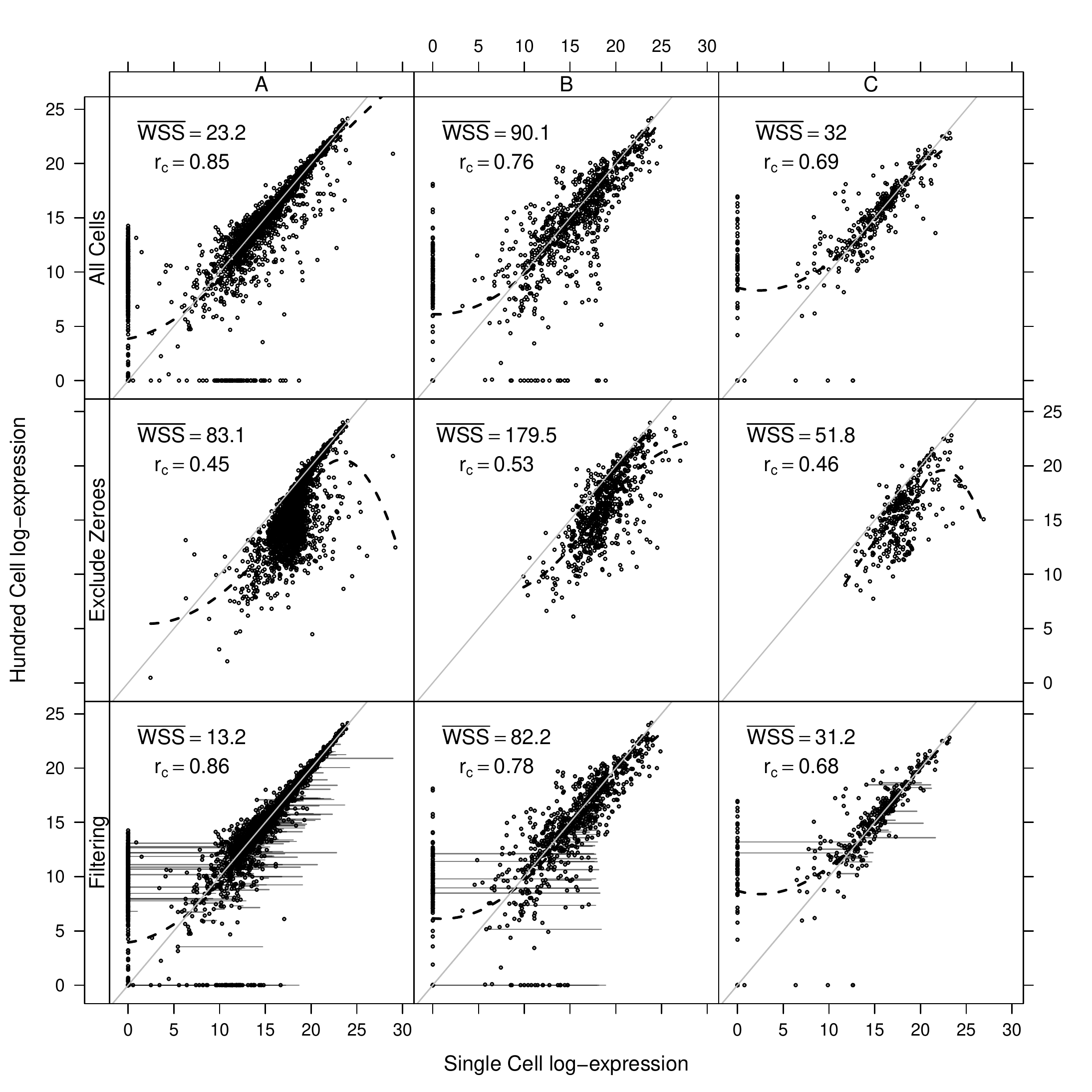}}
\centering
  \caption{Concordance between hundred cell $y^{(100)}/100$ and $y^{(1)}$, the \emph{in silico} average of single cell wells for data sets A, B, C.
In the top row, wells with $v_{ij}=0$ are included and treated as exact zeroes.
In the middle row they are excluded, resulting in a clear lack of concordance.
In the final row, wells are filtered as per section \ref{sec:qual-contr-filt}.
Dark, thin lines show the initial location of a gene before filtering and connect to the location of the gene after filtering.
In each panel, $r_c$, the concordance correlation coefficient and
$\WSS$, the average weighted squared deviation of expression
measurements is printed. The dotted black line shows a lowess fit
throught he data. In all cases, the expression values are transformed using a shifted log-transformation ($\log_2(y+1)$). As such a graphed value of zero corresponds to a zero expression value (i.e. $y=0$).}
\label{fig:etconcord}
\end{figure*}

%A dashed line plots the loess-smoothed regression.
The concordance here is assessed both visually by plotting $\log_2(y_{jk}^{(1)}+1)$ \textit{vs.} $\log_2(y_{jk}^{(100)}+1)$ (Figure \ref{fig:etconcord})
and by calculating the concordance correlation coefficient ($r_c$) between the two variables, which is often used to quantify reproducibility~\citep{Lin1989Concordance}. 
The shifted log transformation allows visualization of both the discrete and continuous components while being on the $\et$ scale.

We first use this concordance experiment to test whether wells that do not cross the fluorescence threshold after $c_{\mathrm{max}}$ should be treated as exact zeroes or missing values.
If we suppose that $v_{ij}=0$ implies an assay failure and the measurement should be discarded, we would simply compute the single cell average over expressed cells, \textit{i.e.}
$y_j^{(1)} = \sum_{i}y_{ij}v_{ij}/\sum_iv_{ij}$. Figure~\ref{fig:etconcord} demonstrates good concordance between the hundred-cell and single-cell experiments when the undetected genes are treated as zeros. However, this is not the case when the zeros are treated as missing values.

% \subsubsection{Normalizing with housekeeping genes does not increase replicate concordance}
% On the other hand, we do not find evidence that the use of traditional housekeeping genes significantly improves concordance between experiments.
% One challenge to the use of housekeeping genes is that they are sometimes not expressed in an otherwise apparently-successful reaction.
% Another challenge is the poor concordance between  housekeeping genes.
% Furthermore, it is known that many putative housekeeping genes are not stably expressed across biological conditions. %cite something

% The housekeeper GAPDH was measured in all experiments and was highly expressed (expression frequency $\pi=$.9-.98).
% Let $\delta_i \equiv \et_{i, \mbox{\gapdh}} - \mu_{\mbox{\gapdh}}$ be the difference between the mean of GAPDH and its observed value in well $i$.
% If $\et_{i, \mbox{\gapdh}} =0$, then we set $\delta_i =0$.
% Then the housekeeper normalized value of gene $j$ in well $i$ is defined as
% \[
%  \et^\prime_{i,j} \equiv \et_{i,j} - \delta_{i},
% \]
% if $\et_{i,j}$ is expressed, and zero otherwise.
% This is analogous to the delta-Ct method used in \rtqPCR.

% In the third row of figure \ref{fig:etconcord}, the hundred cell-single cell concordance is plotted for $\et^\prime$ normalized wells.
% Normalization does not significantly change the concordance.
% In fact, we find that outlying cells (which would be subject to our filtering criteria) drive most of the apparent co-expression between housekeepers.
\subsection{Filtering outlying cells}
In addition to the concordance measure $r_c$, we use another
goodness-of-fit measure to optimize our filtering parameters $t_z$,
$t_\zeta$ %and $noutlier$,
defined by,
\begin{equation}
\WSS = \sum_{j,k} n_{jk} \left (\log_2(y_{jk}^{(1)}+1) - \log_2(y_{jk}^{(100)}+1)\right)^2/JK
\label{eq:wss}
\end{equation}
where $n_{jk}=\sum_iv_{ijk}$ is the number of positive wells for gene
$j$ in unit $k$ in the single-cell experiments.
For a particular gene and unit, the WSS decreases as we lower the
filtering threshold and extreme values are filtered.  Eventually, so
many cells are removed that there is zero expression (and a large
deviance) for the
\textit{in-silico} estimate. Thus we wish to find a set of values for
the filtering parameters that would lead to the lowest WSS measure
across the three data sets used here. The addition of the scaling
factor $n_{jk}$ gives higher weight to combinations with more \emph{ex
  ante} positive observations, so that the contribution to the sum
of squares would be smaller in gene$\times$unit combinations that have fewer expressed cells.
The factor $n_{jk}$ can also be interpreted as the scaling factor for the variance of the mean over positive observations. Finally, the $\overline{\mathrm{WSS}}$ is computed
on the $\log_2(y+1)$ scale to reduce the effect of extreme outliers.

When hundred-cell aggregates are available, one can optimize the
filter parameters $t_\Uz, t_\fz$ %and \emph{noutlier}
by minimizing the WSS over possible combinations.
In our case, we found that setting $t_\Uz=9, t_\fz=9$% and
                                % \emph{noutlier}$=2$
achieves the best reduction in $\WSS$ across the three data-sets explored here (Supplementary Figure~2, and Supplementary Table~1).
Using these values, our filtering criteria moderately improves the concordance between the single-cell and hundred-cell experiments in two of the data sets but dramatically improves (decreases) the weighted sum of squares.
We see that the average per-unit $\et$ of multiple genes are moved towards the diagonal.

\subsection{Normalization and housekeeping genes}
Other authors have noted that ``the gene transcript number is ideally standardized to the number of cells'' \citep{Vandesompele2002Accurate}, which is the case with gene expression from sorted cells.
So it is not entirely a surprise that we find little evidence for housekeeping genes providing useful normalization here.
For a housekeeper to have good validity, it should have high cross correlation with other housekeeping genes.
This is not the case for housekeepers GAPDH and POLR2A, which in data set A, in linear regression have an $R^2=.027$.
In Supplementary Figure 3, we observe in scatter plots of housekeepers' $\et$ that the correlation drops even further (to an $R^2 = .017$) after filtering outlying cells (see previous section).
Since the correlation between housekeepers is present primarily in cells we suspect suffered from technical error, we find little utility in normalization schemes.
In fact, the use of housekeeping genes for normalization could even result in masking cellular artifacts that should be filtered.

\subsection{An efficient test of differential expression for single-cells}
\begin{figure}[!tpb]
  \centering
\includegraphics[width=86mm]{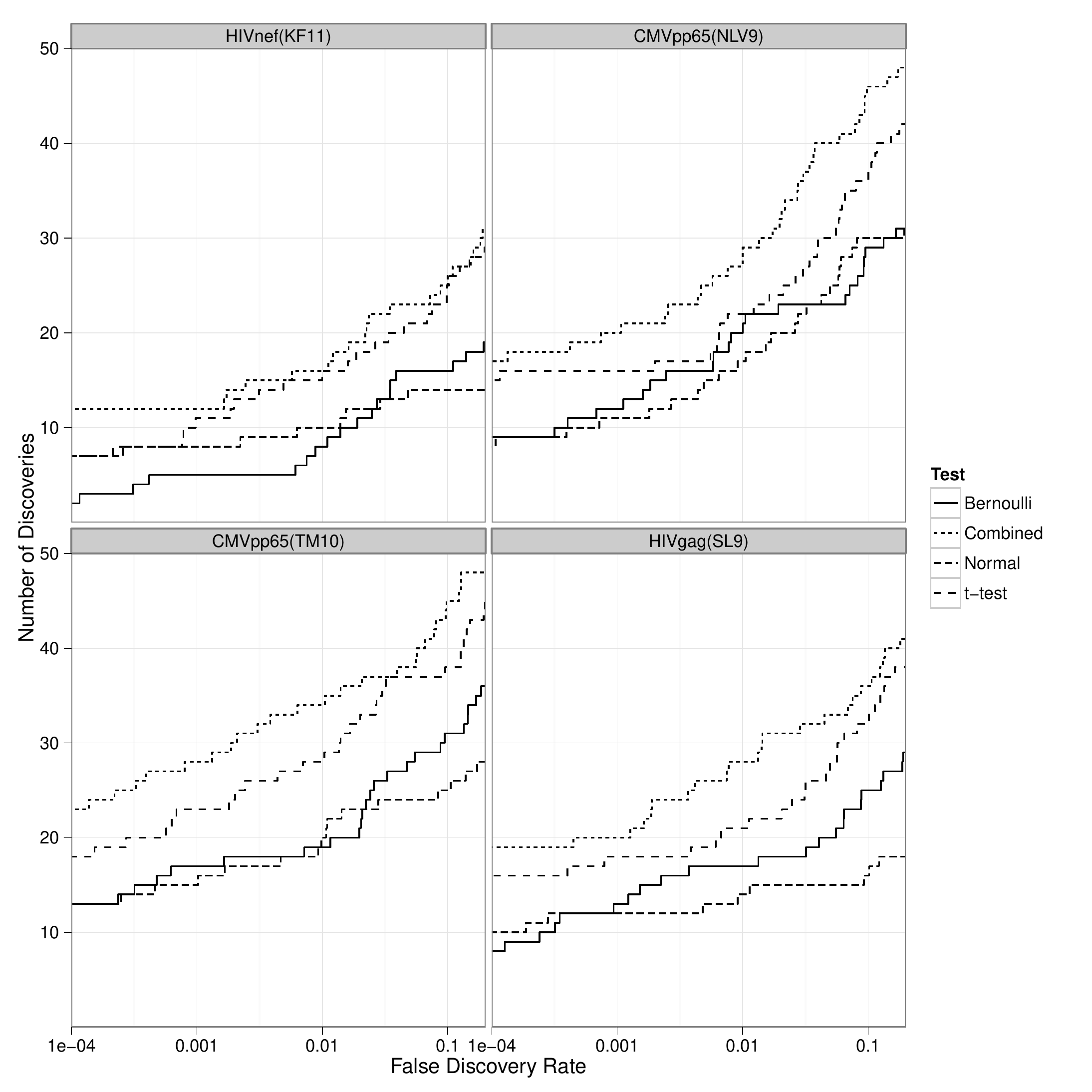}
  \caption{Number of discoveries (genes $\times$ units) versus False Discovery Rate, by treatment, data set A.
The combined likelihood ratio test is compared to a Bernoulli or normal-theory only likelihood ratio test, as well as a t-test of the raw expression values ($2^{et}$ scale), including zero measurements. }
  \label{fig:tests}
\end{figure}
In data set A, approximately 90 cells in each of 16 subjects were measured in unstimulated and stimulated states (see \ref{sec:datasets}). This permits conducting a test for each gene in each subject for differences in $\pi$ and $\mu$, as described in section \ref{sec:test-et-diff}.  We plot the number of discoveries at various false discovery rates (FDR) in Figure \ref{fig:tests}.  The combined likelihood test produces the greatest number of discoveries over a wide range of FDR. For example, at an FDR of 1\%, our combined test could detect more than 20 additional gene$\times$unit changes across the four stimulations.

\begin{figure}[!tbp]
  \centering
  \includegraphics[width=86mm]{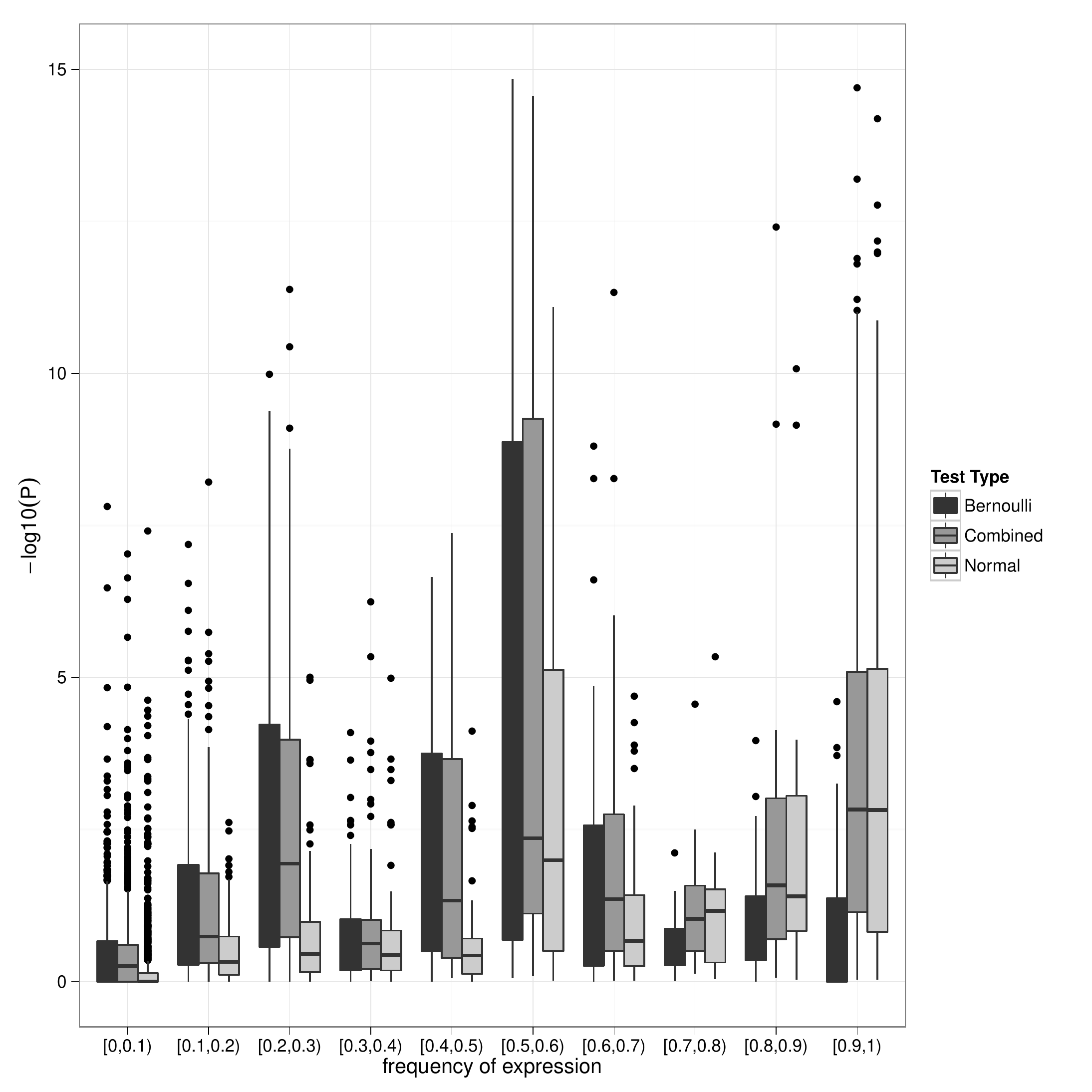}
  \caption{$-\log_{10} P$ of tests (genes $\times$ units) versus frequencies of expression $\pi$ of the genes.  The Bernoulli, normal-theory and combined likelihood ratio tests are plotted.}
  \label{fig:testboxplot}
\end{figure}

Another feature of the combined LRT is its robustness to background gene frequency $\pi_j$.
Of course, if $\pi_j \approx 0$ on average, then any test will be unpowered to detect group differences.
But using only the continuous components amounts to ``throwing away'' data for genes with intermediate $\pi_j$.
And similarly, using only the dichotomous component results in a test insensitive to differences in $\mu_j$ in frequently expressed genes.
This robustness to the $\pi_j$ spectrum is shown in Figure \ref{fig:testboxplot} in which $-\log_{10}$ p-values are shown for the Bernoulli, normal and combined LRTs versus frequency of $\pi_j$.

\begin{figure}[tbp]
  \centering
  \includegraphics[width=86mm]{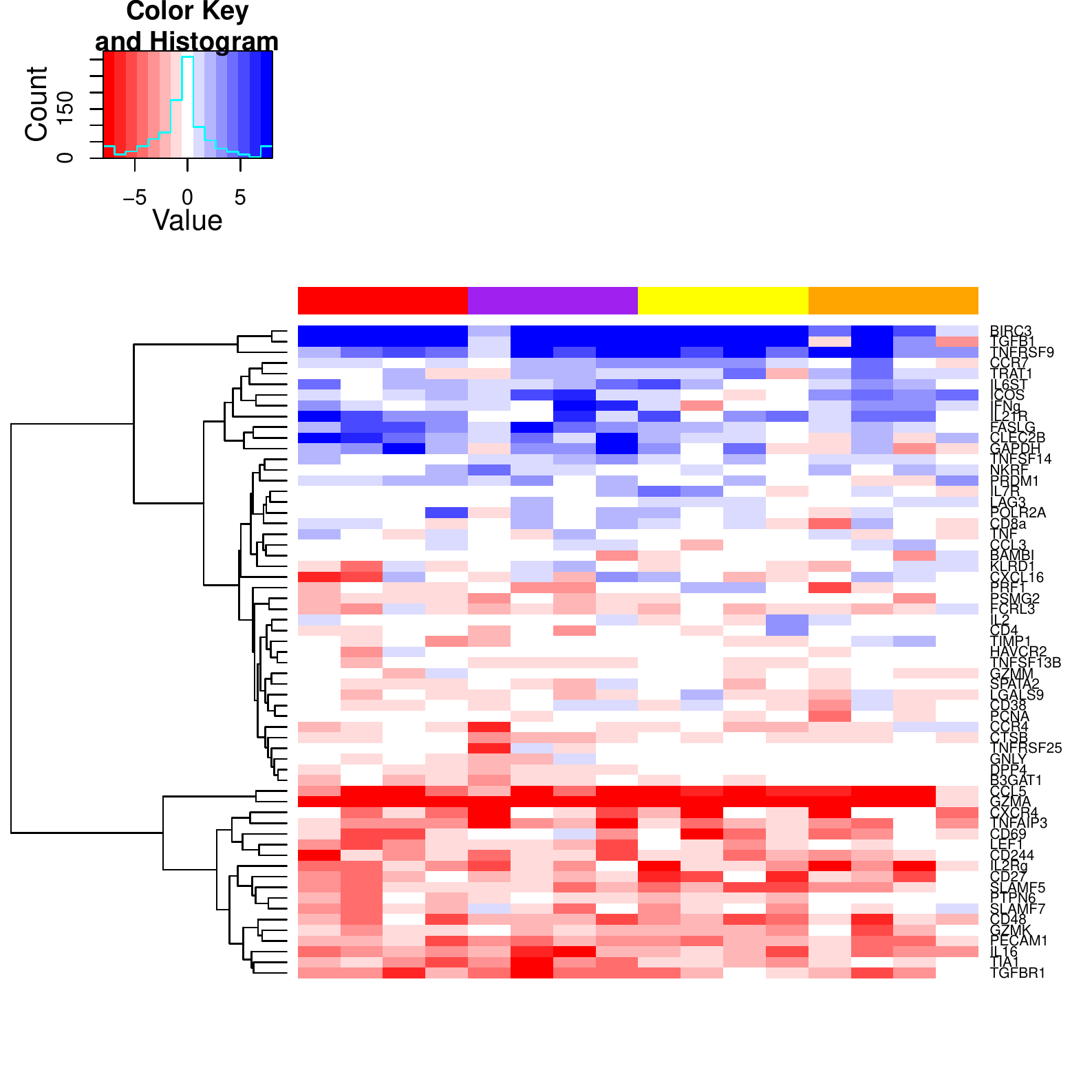}
  \caption{Heatmap of signed $\log_{10} p$ for selected genes (rows) and all individuals (columns).  The color above each column indicates the stimulus applied to the cells.  Red and purple are two CMV peptide pools; yellow and orange are two HIV peptide pools. }
  \label{fig:heatmap}
\end{figure}

A total of 65 genes were detected at an FDR of 1\% in at least one individual.
We define $p^\star =  -\mbox{sign} (\mu_1 - \mu_0) \cdot \log_{10} p$ as the negative $\log_{10}$ p-value times an indicator variable which is positive when stimulated groups have greater expression, and negative otherwise.
Figure~\ref{fig:heatmap} plots a heatmap of signed $\log_{10}$ p-values.
The selected genes are in clustered rows; individuals are in columns arranged by stimulations.
The color above each column indicates which antigen stimulation the individual received.
From this, it is clear that genes cluster into up-regulated and down-regulated modules and that there is much individual variability in response.
%\subsection{Effect of Filtering on Differential Expression}

%\section{Discussion}
%\subsection{Stimulation specific expression patterns}

\section{Conclusion}
Current approaches for analysis of single-cell assays have incompletely utilized the salient features of the experiment, and the resulting inference can be sub-optimal.
In this paper, we have presented a framework for data exploration, quality control and testing for differential expression using single-cell data.
Our comparison of 100-cell  and single-cell measurements shows that undetected genes in an assay should be treated as effective ``zeroes.''
Both the discrete, zero-inflated portion and continuous portion of single-cell expression data are meaningful for detecting outliers.
Moreover, differences in either could be of biological interest, so it is desirable to combine evidence from both to detect changes in expression.
Our likelihood ratio test allows just that.

Although we have suggested default parameters for the filtering of outliers, informed from several data sets, our defaults are likely conservative.
They are 3-4 times larger than the most substantial difference in expression between experimental groups we observed.
Therefore, scientists may wish to tune these parameters based on their own relative importance of eliminating potential technical error versus missing biological heterogeneity.
Acquiring forms of ground-truth besides ``bulk'' experiments (in our case, 100-cell aggregates) could allow forming tighter bounds.

Further work, incorporating a mixed-effects model to our likelihood ratio test, could extend its applicability.
The test outlined in this paper may not be appropriate in cases where traits of interest are not blocked within individuals (\textit{e.g.}, comparing between phenotypes like HIV+ \textit{vs.} HIV-).
In this case, one wishes to identify gene expression changes across groups, in spite of high individual-to-individual heterogeneity.
By modeling the mean and proportion of expression as common across
groups and adding specific random effects for between-individual
variability, our model could be extended to address such experimental questions as well.

Single-cell gene expressions assays have already been shown to be useful in multiple studies and will become even more routine once sequencing at the single-cell level becomes practical~\citep{Varadarajan2011Highthroughput,Ramskold2012Fulllength}.
As a consequence, the development of effective statistical methods to analyze such data is becoming increasingly important.
This paper offers a coherent framework for researchers using this nascent technology.

\section*{Acknowledgments}
\paragraph{Funding:}
This work was supported by the Intramural Research Program of the National Institute of Allergy and Infectious Diseases, National Institute of Health, and the Collaboration for AIDS Vaccine Discovery [\#38650]; National Institute of Health [U19 AI089986-01, R01 EB008400 to RG, GF and AM]; and the Bill \& Melinda Gates Foundation [\#OPP1032325].

\bibliographystyle{natbib}
\bibliography{amcdavid-eda_template}

\end{document}